\documentclass[a4paper, 11pt, fleqn]{article}
\usepackage{longtable}
\usepackage{graphicx}
\usepackage{amssymb,amsmath,epsfig,array}

\begin{document}


\title{\bf{Multicolor photometry and spectroscopy of the yellow
supergiant with dust envelope HD~179821=V1427~Aquilae \ }}

\author{N.P. Ikonnikova\footnote{E-mail: ikonnikova@gmail.com},
\fbox{O.G. Taranova}, V.P. Arkhipova, \\
G.V. Komissarova, V.I. Shenavrin, V.F. Esipov, M.A. Burlak,
V.G. Metlov}

\date{\it{Sternberg Astronomical Institute,\\ Moscow State University (SAI MSU), Universitetskii pr. 13, Moscow, 119992 Russia}}

\renewcommand{\abstractname}{ }

\maketitle

\begin{abstract}

We present the results of multicolor ($UBVJHKLM$) photometry
(2009-2017) and low-resolution spectroscopy (2016-2017) of the
semi-regular variable V1427~Aql=HD~179821, a yellow supergiant
with gas-dust envelope. The star displays low-amplitude ($\Delta V
<0.^{m}2$) semi-periodic brightness variation superimposed on a
long-term trend. The light curve shape and timescale change from
cycle to cycle. There are temperature variations characteristic
for pulsations, and brightness oscillations with no significant
change of color are also observed. The $UBV$ data for the
2009-2011 interval are well reproduced by a superposition of two
periodic components with $P=170^{d}$ and $P=141^{d}$ (or
$P=217^{d}$ -- the one year alias of $P=141^{d}$). The variation
became less regular after 2011, the timescale increased and
exceeded 250$^{d}$. An usual photometric behavior was seen in 2015
when the star brightness increased by $0.^{m}25$ in the $V$ filter
in 130~days and reached the maximum value ever observed in the
course of our monitoring since 1990. In 2009-2016 the annual
average brightness monotonically increased in $V$, $J$, $K$,
whereas it decreased in $U$ and $B$. The annual average $U-B$,
$B-V$, and $J-K$ colors grew, the star was getting redder. The
cooling and expanding of the star photosphere along with the
increasing of luminosity may explain the long-term trend in
brightness and colors. Based on our photometric data we suppose
that the photosphere temperature decreased by $\sim$400~K in the
2008-2016 interval, the radius increased by $\sim$24~\%, and the
luminosity grew by $\sim$19~\%. We review the change of annual
average photometric data for almost 30 years of observations.
Low-resolution spectra in the $\lambda4000-9000$~\AA\ wavelength
range obtained in 2016-2017 indicate significant changes in the
spectrum of V1427~Aql as compared with the 1994-2008 interval,
i.e. the BaII and near-infrared CaII triplet absorptions have
turned stronger while the OI $\lambda$7771-4 triplet blend has
weakened that points out the decrease of temperature in the region
where the absorptions are formed. The evolutionary stage of the
star is discussed. We also compare V1427~Aql with post-AGB stars
and yellow hypergiants.

{Keywords: \it{post-AGB, post-RSG, yellow hypergiants, spectral
and photometric observations, brightness variability, evolution}}

\end{abstract}

\newpage

\section*{INTRODUCTION}

The evolutionary status of the yellow supergiant with infrared
(IR) excess HD~179821 (BD--00$^{\circ}$3679 = SAO~124414 =
IRAS~19114+0002 = V1427~Aql) is not entirely clear. No doubt it is
an evolved star as implied by the presence of a spherical gas-dust
circumstellar envelope formed through mass loss episodes during
earlier stages of evolution. But the initial mass of HD~179821 is
still under the question. Some data point to intermediate mass
($M_{ZAMS}=1\div8M_{\odot}$), and then HD~179821 is a post-AGB
supergiant and a planetary nebula progenitor (Pottasch and
Parthasarathy 1989; Hrivnak et al. 1989; Ferguson and Ueta 2010);
other data favor the idea that HD~179821 is a massive star
($M_{ZAMS}>20~M_{\odot}$) which has evolved off the Red Supergiant
(RSG) and later can undergo a Supernova II outburst, lose its
envelope, and become a neutron star (Jura et al. 2001; Oudmaijer
et al. 2009; \c{S}ahin et al. 2016). Both hypotheses were
discussed in detail in Reddy and Hrivnak (1999), Josselin and
Lebre (2001), Coroller et al. (2003), Arkhipova et al. (2009).

Photometric instability is an important feature of HD~179821. We
discovered its variability on the basis of our photometric
observations in 1990-1992 (Arkhipova et al. 1993). In the General
Catalogue of Variable Stars, the object was designated as
V1427~Aql and associated with semi-regular SRd-type variables
(Samus 2017). Further investigations carried out in 1993-1999
(Arkhipova et al. 2001), in 1994-2000 (Hrivnak 2001), in 1999-2000
(Coroller et al. 2003), and in 2000-2008 (Arkhipova et al.2009)
revealed low-amplitude ($\Delta V=0.^{m}1-0.^{m}2$) brightness
oscillations with cycle lengths of nearly 200~d and also a general
change in brightness and colors. The stars at the late stages of
evolution are known to experience some instability which differs
for post-AGB supergiants and yellow hypergiants, the latter having
evolved off the RSG. Therefore knowing photometric variability
peculiarities is of particular importance for stars of uncertain
evolutionary status.

In this work, we present new multicolor photometric data for
V1427~Aql obtained in 2009-2017 and analyze the overall change in
brightness and colors over the whole period of our photometric
observations since 1990. We also adduce low-resolution spectral
observations obtained in 2016-2017 and compare them with earlier
data.

\section*{OBSERVATIONS}

\subsection*{$UBV$-photometry}

We have been carrying out photometric observations of V1427~Aql
since 1990. The results of $UBV$ light curve study over the
1990-2008 interval were reported in Arkhipova et al. (1993, 2001,
2009).

After 2008 we continued to monitor V1427~Aql at the Crimean SAI
MSU Station with the 60-cm Zeiss reflector equipped with the
photon counting $UBV$-photometer (Lyutyi 1971). The diameter of
the chosen aperture was 27$''$. The star SAO~124412 (sp=G8IV) with
$U$=9.$^{m}$87, $B$=9.$^{m}$56, $V$=8.$^{m}$77 was used as a
comparison star. We estimate the average uncertainty to be about
$0.^{m}$01. Table~1 presents $UBV$-magnitudes for V1427~Aql in
2009-2017.

\subsection*{IR-photometry}

IR-photometric observations were made at the Crimean SAI MSU
Station with the 125-cm telescope. The photometer with a liquid
nitrogen cooled photovoltaic indium antimonide (InSn) detector
(Shenavrin et al. 2011) was installed at the Cassegrain focus, the
output aperture was $\sim$ 12$''$. The star BS~7377 (Sp=F0IV) was
used as a photometric standard, and its magnitudes were taken from
the catalog of Johnson et al. (1966): $J=2.^{m}75$, $H=2.^{m}61$,
$K=2.^{m}57$, $L=2.^{m}54$, $M=2.^{m}59$ . Our photometric errors
did not exceed $0.^{m}05$ for the $M$-band and $0.^{m}02$ for the
others. In an earlier paper (Arkhipova et al. 2009), we review the
IR photometry made in 1992-2000 and in 2008 (two nights) and also
the data obtained by other researchers before 1992. Table~2
present new $JHKLM$-magnitudes for HD~179821 in 2009-2017.

\subsection*{Spectral observations}

We have been making low-resolution spectral observations of
V1427~Aql since 1994. The review of spectral data obtained in
1994-2008 can be found in Arkhipova et al. (2009). In this work,
we present the recent observations made in 2016-2017. Spectra with
a wavelength coverage of $\lambda$4000-9000~\AA\ were obtained at
the Crimean SAI MSU Station with the 125-cm reflector and the fast
A-spectrograph. An SBIG ST-402 CCD detector was employed providing
a spectral resolution of $\sim$ 2.2 \AA\ per pixel.

The log of observations is presented in Table~3, where we list the
observing date and the standard star used for calibration. On
October 25, 2016, the standard star was not observed because the
weather got worse.

The spectra were processed with the use of the CCDOPS standard
program and the SPE program (Sergeev and Heisberger 1993). The
data were flux-calibrated using spectrophotometric standard stars,
the spectral energy distribution for them in the
$\lambda$4000-7650 \AA\ wavelength range was taken from the
spectrophotometric catalog of Glushneva et al. (1998) and extended
to $\lambda$9000 \AA\ according to the stellar spectral flux
library of Pickles (1998).

\section*{ANALYSIS OF PHOTOMETRIC AND SPECTRAL VARIABILITY OF V1427~AQL}

\subsection*{Photometric variability of V1427~Aql in 2009-2017.}

In Fig.~1 we present the $UBV$ light and $U-B$, $B-V$, $U-V$ color
curves for V1427~Aql observed in Crimea in 2009-2017. During that
time the object's behavior was not strictly homogeneous.


\begin{figure}

\includegraphics[scale=1.85]{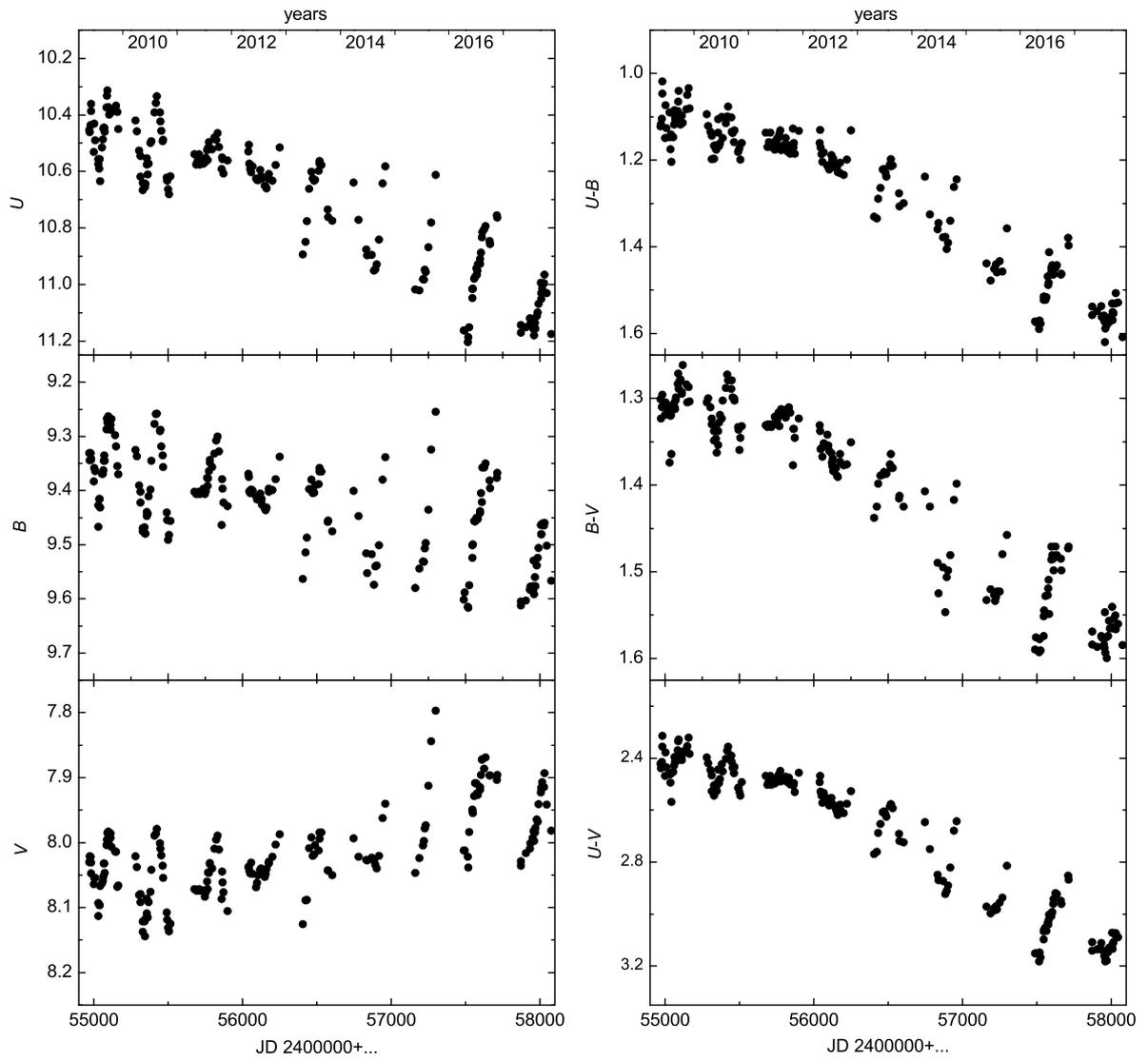}
\caption{Light and color curves of V1427~Aql in 2009-2017.}
\end{figure}


In 2009-2010 (JD=24~54970-55516) V1427~Aql displayed a regular
cyclical variation in brightness. Colors varied strictly in phase
with brightness: the object was bluer when brighter and redder
when fainter. In 2010, when the pulsation pattern appeared the
most distinct, the full range in brightness was of $\Delta
V=0.^{m}18$ and the $B-V$ color varied over a range of
$0.^{m}58-0.^{m}66$.

A smaller brightness range was present in the 2011 observations
with a peak-to-peak variation of nearly $0.^{m}1$ in $UBV$.
Brightness variations were not accompanied by significant change
in color.

Brightness variations turned to be less regular in 2012-2014
(JD=24~56040-56605) with maximum $V$ amplitude of $0.^{m}15$, the
timescale of brightness variability increased and became
comparable to the duration of an observational season.

In 2015 we performed $UBV$-observations of V1427~Aql during 138~d,
from May 19 till October 3 (JD=24~57162-57299). In the course of
this time, the star brightened quickly by $\Delta V\sim0.^{m}25$,
$\Delta B\sim0.^{m}33$ and $\Delta U\sim0.^{m}40$. By JD=24~57299
the $V$ brightness reached the value of $7.^{m}85$ that appeared
the maximum brightness since 1990, over the whole interval we had
observed V1427~Aql. Unfortunately, the estimate obtained that
night was the last in 2015, so we could not fix the moment of
maximum in that cycle. The evolution of the stellar colors was
unusual: the brightness increasing from $V=8.^{m}05$ to
$V=7.^{m}91$, the $B-V$ and $U-B$ colors hardly varied indicating
that the luminosity growth was due to the increase in the star
radius but not to the rise of the temperature; then the star
became a little bluer.

The next observational season for V1427~Aql started on April 7,
2016 (JD=24~57486) and lasted for 228~days and during this time
the stellar brightness increased systematically in all bands
displaying local extremes in the $B$ and $U$ light curves: minimum
brightness at JD=24~57517 and maximum brightness at
JD$\sim$24~57630. The star brightened by $\Delta V\sim0.^{m}15$,
$\Delta B\sim0.^{m}27$, $\Delta U\sim0.^{m}39$ having turned bluer
by $\Delta(B-V)\sim0.^{m}10$ and $\Delta(U-B)\sim0.^{m}15$.

The stellar brightness was increasing for 146~days in 2017,
beginning April 26 (JD=24~57870), reached its maximum near
September 18 (JD=24~58015), and then the star started getting
fainter. The amplitude of light variation was up to $0.^{m}13$.

Fig.~2 plots the color-color and the color-magnitude diagrams to
demonstrate the difference in color-magnitude relation for various
observational seasons.

Apart from regular oscillations of brightness and color which are
more or less regular and correspond to pulsations and maybe some
other type of atmospheric instability the star displayed a general
trend in brightness and color over the observing interval: on
average the $V$ brightness increased from season to season with
the $B$ and $U$ brightness getting fainter and the $U-B$ and $B-V$
colors getting redder.


\begin{figure}
\includegraphics[scale=1.2]{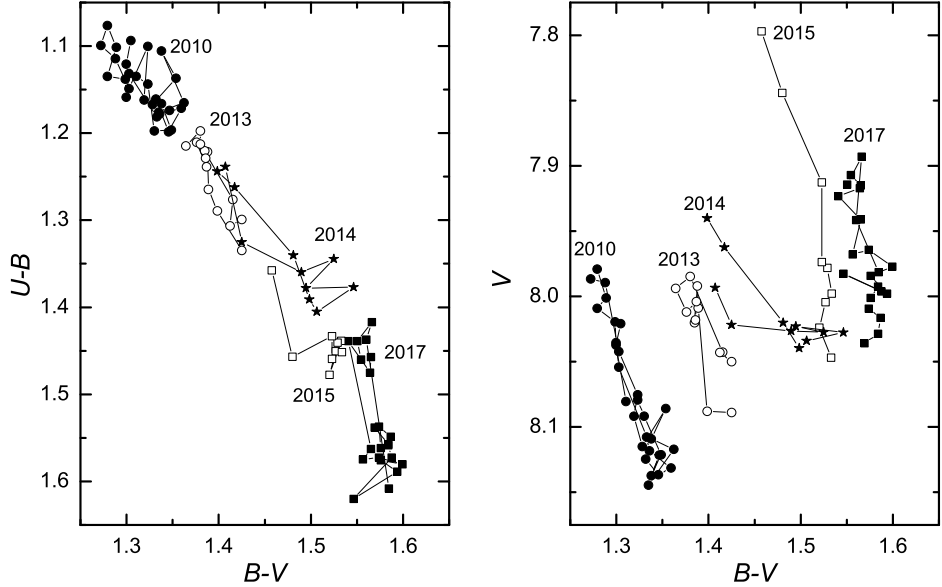}

\caption{Color-color and color-magnitude diagrams for V1427~Aql
covering five observing seasons. Numbers indicate years.}
\end{figure}


In the near-IR range, the $J$ (1.25~$\mu$m) and $K$ (2.2~$\mu$m)
observations are the most numerous for the 2009-2017 interval.
Fig.~3 shows $J$, $K$, $L$ and $M$ light and $J-K$, $H-K$, $L-M$
color curves. Stellar pulsations are responsible for brightness
variations within each season with the amplitude not exceeding
$\sim0.^{m}1$ and being largest in the $J$ band. The $JHKL$
brightness appeared to increase systematically from season to
season in 2009-2015. By 2015 the star showed an increase in
brightness of $0.^{m}4$ in $J$ while colors turned redder, for
example, the $J-K$ color increased by $\sim0.^{m}1$. It's worth
mentioning that a significant excess of $M$ light was observed in
2015.

In the section "Secular trend in brightness and color"\ we discuss
the gradual change of seasonal average photometric parameters of
the star.


\begin{figure}
\includegraphics[scale=1.85]{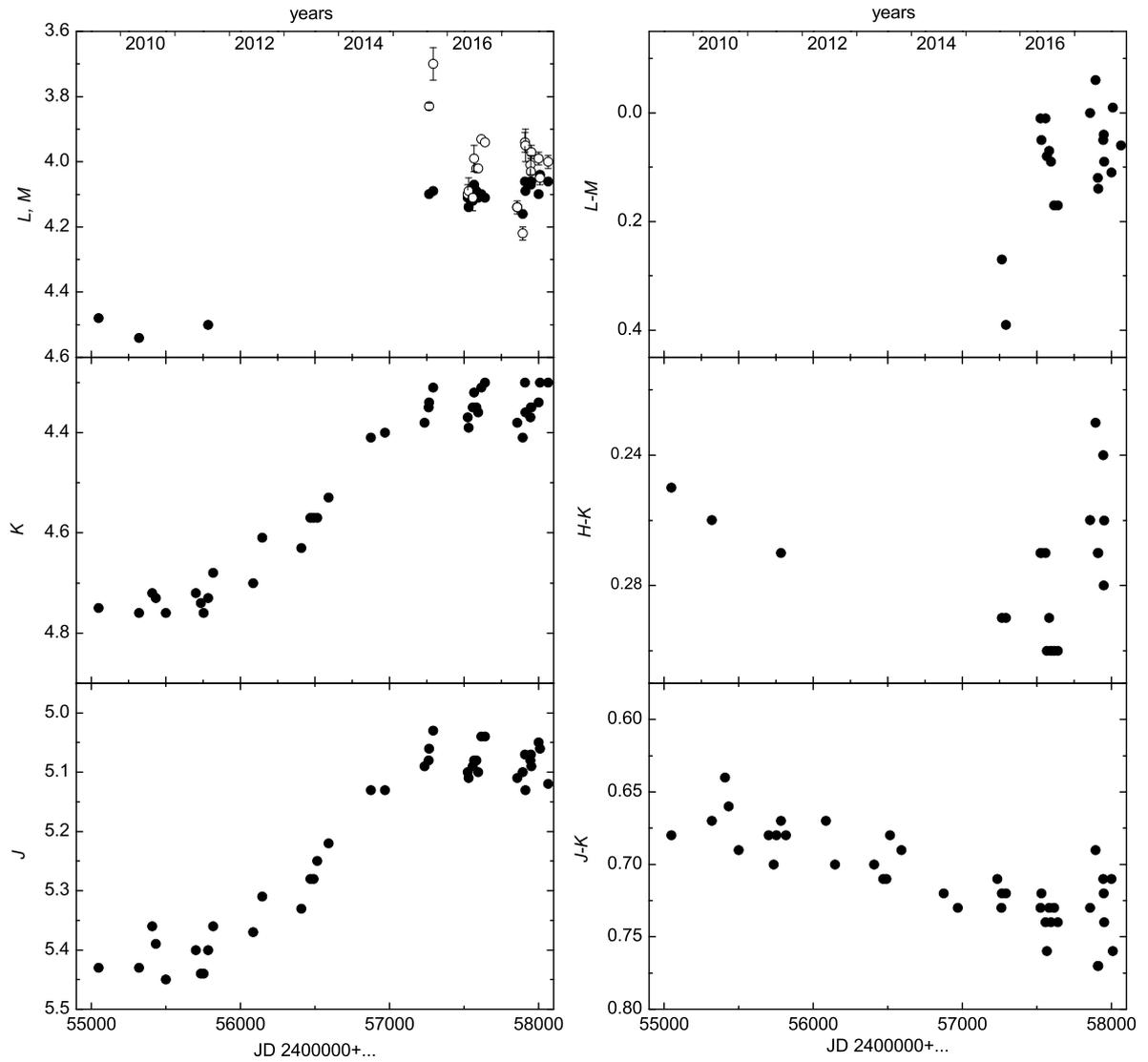}
\caption{Light and color curves of V1427~Aql in near-IR range in
2009-2017. In the upper left panel points represent $L$
magnitudes, open circles -- $M$ magnitudes.}
\end{figure}


\subsection*{Period analysis}

Fig.~4 shows the entire $V$ light curve observed in 1990-2017. As
one can see the star displays semi-regular brightness variation
with changing amplitude that is typical for oscillations with
multiple periods.


\begin{figure}
\includegraphics[scale=0.3]{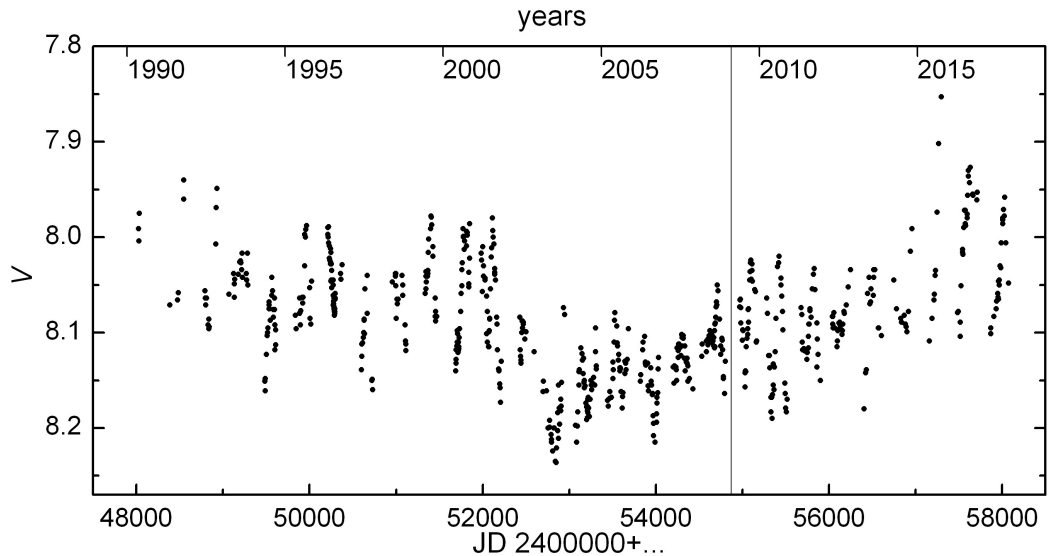}
\caption{$V$-band light curve for V1427~Aql observed in 1990-2017.
More recent data obtained in 2009-2017 are located to the right
from the vertical line.}
\end{figure}


Search for periodicity for V1427~Aql was carried out by Arkhipova
et al. (2001, 2009) on the basis of photometric data obtained in
1990-1999 and in 2000-2008 respectively and also by Coroller et al
(2003) on the basis of their own observations made in 1999 and
2000 with the data of Arkhipova et al. (2001) and Hrivnak (2001)
engaged.

Using $UBV$ observations of V1427~Aql made in 1994-1999 (Arkhipova
et al. 2001), we found two periods of $P_{0}=205^{d}$ and
$P_{1}=142^{d}$ and suggested that they correspond to the
fundamental mode and the first overtone of pulsation. Coroller et
al. (2003) got similar values.

Arkhipova et al. (2009) pointed out a possible change of pulsation
period and phase for V1427~Aql after the year 2000. We found two
periods $P_{0}=196\pm5^{d}$ and $P_{1}=128\pm5^{d}$ for the
interval 2001-2008. These periods are the one year alias of each
other. It turned difficult to determine which of them is a true
period of pulsations and which is a spurious one. The observations
made in 2001, 2004 and 2005, when the pulsation pattern was
distinct, agree better with $P_{1}=128\pm5^{d}$ while the
timescale of brightness change was closer to 200$^{d}$ in 2002,
2003, 2006 and 2007.

In this work, we present frequency analysis basing on the new
$UBV$-data acquired in 2009-2017.

To search for periods, we used the program PERIOD04 (Lenz and
Breger 2005) and also the program EFFECT created by V.P.~Goranskij
which applies a Fourier transform for arbitrary data spacing
(Deeming 1975).

The frequency spectrum derived from the $UBV$ data for the
2009-2011 interval shows a clear peak at $P_{1}=170\pm5^{d}$, and
with this peak removed, two other periods dominate the residual
spectrum: $P_{2}=141\pm5^{d}$ and $P_{3}=217\pm5^{d}$. The
frequency spectrum that covers the period range of
$10^{d}-400^{d}$ and includes the 2009-2011 $B$ data is shown in
Fig.~5. Ten estimates of $J$ brightness obtained in the interval
2009-2011 yield two periods of 137$^d$ and 216$^d$ which are the
one year alias of each other and are close to the values $P_{2}$
and $P_{3}$ derived from the $UBV$ data.

In 2012-2017 the period increased and exceeded observing
intervals. The timescale of brightness variability rose up to
260$^d$ in 2012-2014. The frequency analysis reveals a period of
nearly $370^d$ for the 2015-2017 subset of $V$ data. 22 estimates
in the $J$ band made in 2015-2017 yield a period of $175^d$ but
its significance value is not high.

The nature of V1427~Aql semi-periodic variability is not entirely
clear. Along with the cyclic change in brightness which is similar
to pulsation, variations of another type are present perhaps being
a result of variable stellar wind.


\begin{center}
\begin{figure}
\includegraphics[scale=1.6]{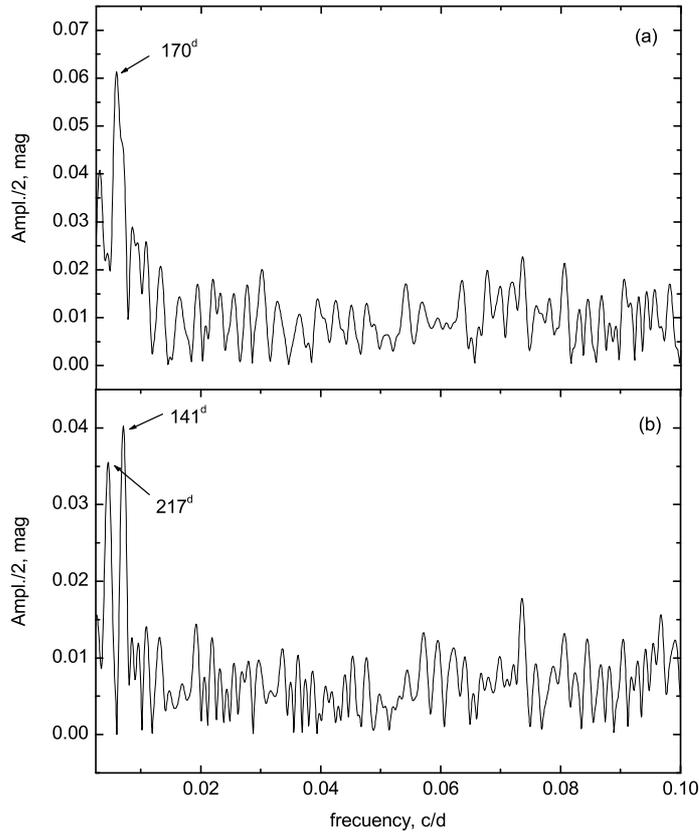}

\caption{The frequency spectrum based on the $B$ data obtained in
2009-2011 for V1427~Aql: a)~the initial spectrum including the
full set of observations; b)~the residual spectrum with the
primary period component removed from the light curve.}
\end{figure}
\end{center}

\subsection*{Secular trend in brightness and color}

Long-term monitoring of V1427~Aql from 1990 till 2017 leads to the
detection of a secular trend in brightness and color. In Fig.~6,
which shows the $V$ light and $U-B$, $B-V$ color curves for the
whole observing interval including data from Arkhipova et al.
(2009) and also the change of $J$ magnitude and $J-K$, $K-L$
colors from 1987 till 2017 including previously published (Hrivnak
et al.1989; Kastner and Weintraub 1995; Arkhipova et al. 2009) and
new data, each point is an annual mean value.


\begin{figure}
\includegraphics[scale=1.85]{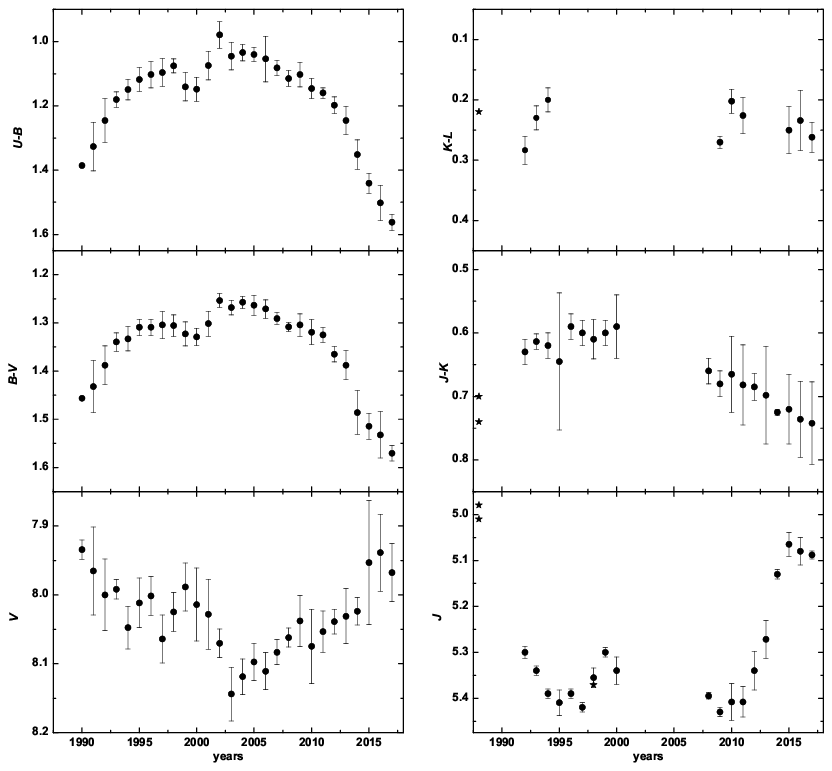}

\caption{Optical and near-IR light and color curves for V1427~Aql
composed of annual mean values. Points represent our data,
asterisks -- original estimates from Hrivnak et al. (1989),
Kastner and Weintraub (1995) and 2MASS catalog. Vertical bars
correspond to the amplitude of variation within a season.}
\end{figure}


The $U-B$ and $B-V$ colors were turning monotonically bluer from
1990 to 1994 with the $V$ brightness decreasing. The $B-V$ color
remained quite constant after 1994, and the $U-B$ color continued
to decline. The 1998-2000 data reveal a tendency for the object to
get redder, then the star turned bluer again in 2001-2002, and in
2002 the $U-B$ and $B-V$ colors reached the values that appeared
minimum for the whole observing interval. Later, after the year
2002, the object was getting redder brightening in the $V$ band.
By 2016 the $V$ brightness reached the 1990 value with the colors
being redder than when our observations started.

Hrivnak et al. (1989) and Kastner and Weintraub (1995) were the
first to obtain near-IR observations of V1427~Aql in 1987-1988. By
1992 when we started monitoring V1427~Aql in the near-IR, the star
was found $\sim0.^{m}3$ fainter in $J$, and the $J-K$ color turned
out bluer. From 1992 till 2000 the $J$ brightness was changing by
not more than $0.^{m}1$ with the $J-K$ color variation being
within the range of $0.^{m}05$. From 2001 till 2007 the star was
not observed in the near-IR. Since 2008 the star was getting
brighter in the $J$, $H$, $K$ and $L$ bands with the $J-H$ and
$J-K$ colors increasing, while from 2015 there appeared a tendency
for the star to get fainter in all filters and for the $J-H$ and
$J-K$ colors to decrease. In the 1-5~$\mu$m wavelength range no
IR-excess was detected for the star, so the energy emitted in this
range is attributed entirely to the stellar photosphere and
determined by its parameters.

Fig.~7 shows the color-color diagram for V1427~Aql with the annual
mean $UBV$ magnitudes used for the interval 1990-2017. As is
apparent from the figure the star was getting bluer in 1990-2002,
and then a systematic reddening began. The fastest change in color
occurred recently in 2013-2017. During this interval, the seasonal
mean colors increased significantly, by $\Delta (U-B)\sim 0.^{m}3$
and $\Delta (B-V)\sim 0.^{m}2$. As we reported in Arkhipova et al.
(2009) the long-term trend in colors coincides with the $B-V$ and
$U-B$ colors varying in case of pulsations. That is why we
consider the main cause of the long-term trend in colors to be the
change of the stellar temperature: growth from 1990 till 2002 and
then systematic decline towards 2017.


\begin{figure}
\includegraphics[scale=1.2]{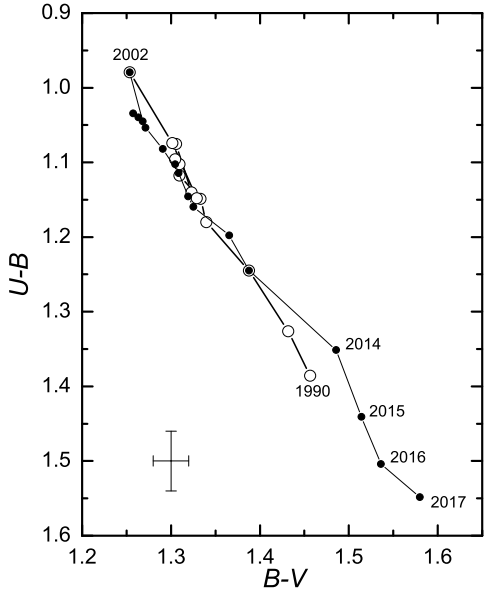}

\caption{$(B-V)-(U-B)$ color-color diagram for the annual mean
magnitudes of V1427~Aql. Open circles represent the 1990-2002
data; points correspond to the 2002-2017 interval. The numbers
indicate years.}
\end{figure}


Fig.~8 shows the color-magnitude relations based on the seasonal
average optical and IR magnitudes for V1427~Aql. The star moves
similarly in both diagrams. After the first observations of
V1427~Aql in 1987-1988, the star appeared to show a monotonic
decrease in the $V$ and $J$ brightness along with getting bluer
that has turned to brightening and reddening in recent years. The
long-term trend in brightness and colors reflects the change in
temperature and luminosity of the star photosphere. In Fig.~8 the
reddening trajectory for the total and selective interstellar
extinction laws $A_{V}=3.1\times E(B-V)$ and $A_{J}=1.50\times
E(J-K)$ adopted from Koornneef (1983) is shown with arrows. If the
circumstellar extinction changed, its range was not larger than
$\Delta E(B-V)=0.^{m}02$.


\begin{figure}
\includegraphics[scale=1.4]{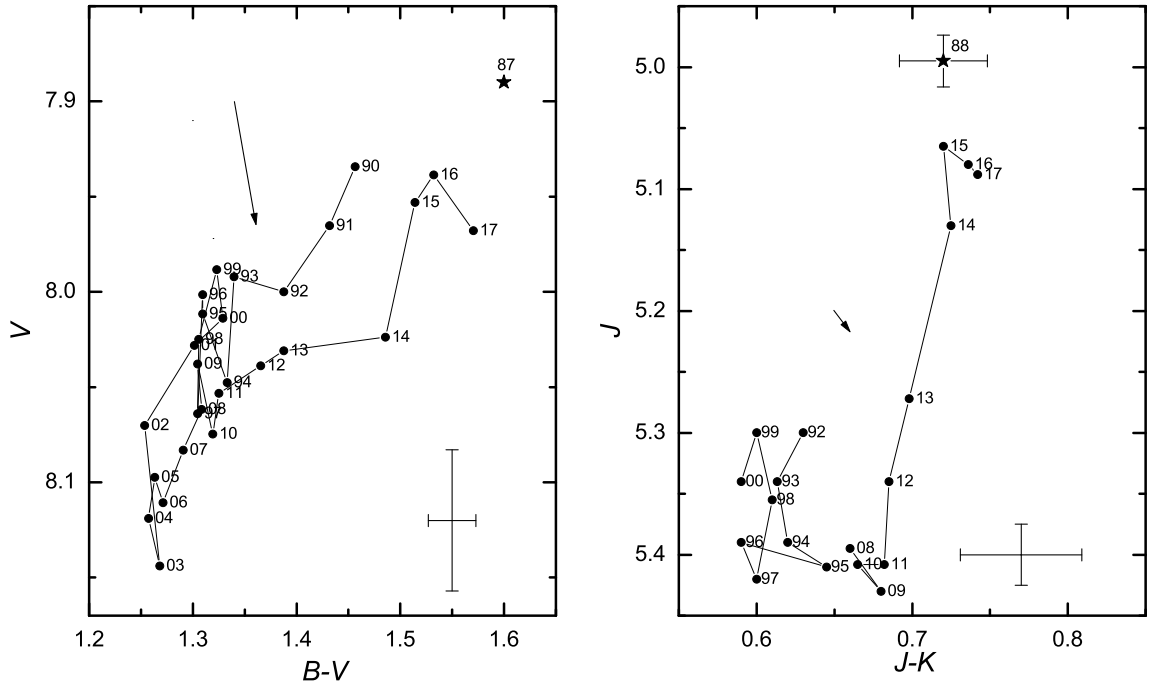}

\caption{The $(B-V)-V$ and $(J-K)-J$ color-magnitude diagrams. The
points represent the seasonal means of our data, the asterisks --
the observations from Hrivnak et al. (1989). The numbers indicate
years. The reddening trajectory for the standard interstellar
extinction law is shown with arrows. The arrows' length
corresponds to the extinction associated with $E(B-V)=0.^{m}02$.}

\end{figure}

\subsection*{Spectral variability of V1427~Aql in 1994-2017.}

Our spectral observations of V1427~Aql obtained in 1994-2008 were
discussed in Arkhipova et al. (2009). In that work, we estimated
the temperature and spectral class of the star having compared it
with other objects, and also reported about the H$\alpha$ line
variability. We did not find other changes in spectrum during the
mentioned interval.

We did not obtain spectra for the star from 2009 till 2015,
recommenced observing V1427~Aql in May 2016 and continued until
October 2017 (see Table~3).

Analysing new observational data we found out that the spectrum of
V1427~Aql had changed a lot between the 1994-2008 interval and the
year 2016. Fig.~9 shows the continuum normalized spectra for
V1427~Aql in the $\lambda$4200--9000 \AA\ wavelength range
obtained on June 7, 2008, and July 27, 2016. The spectral lines to
look at are marked in the plot. The BaII lines are seen to have
got deeper from 2008 to 2016, the absorptions related to the CaII
IR-triplet strengthened. Besides the OI triplet at $\lambda$7771-4
\AA\ and Paschen lines weakened, the G-band to H$\gamma$ line
relation changed.

For the quantitative analysis, we measured the equivalent widths
$EW$ of several  absorptions, namely H$\alpha$, BaII
$\lambda$5857, 6497, FeII $\lambda$6516, the OI $\lambda$7771-4
triplet, P12, P14, and P17 hydrogen lines, the CaII triplet
(CaT=$\lambda$8498+$\lambda$8542+$\lambda$8662), using the spectra
from Arkhipova et al. (2009) and those obtained in 2016 and 2017
and here present them in Table~4. It is worth mentioning that due
to the given spectral resolution the CaII triplet components were
severely blended with the P13, P15 and P16 hydrogen lines. The
estimated error was about 10 \%.

Fig.~10 shows the evolution of the equivalent widths of several lines.


\begin{figure}

\includegraphics[scale=1.0]{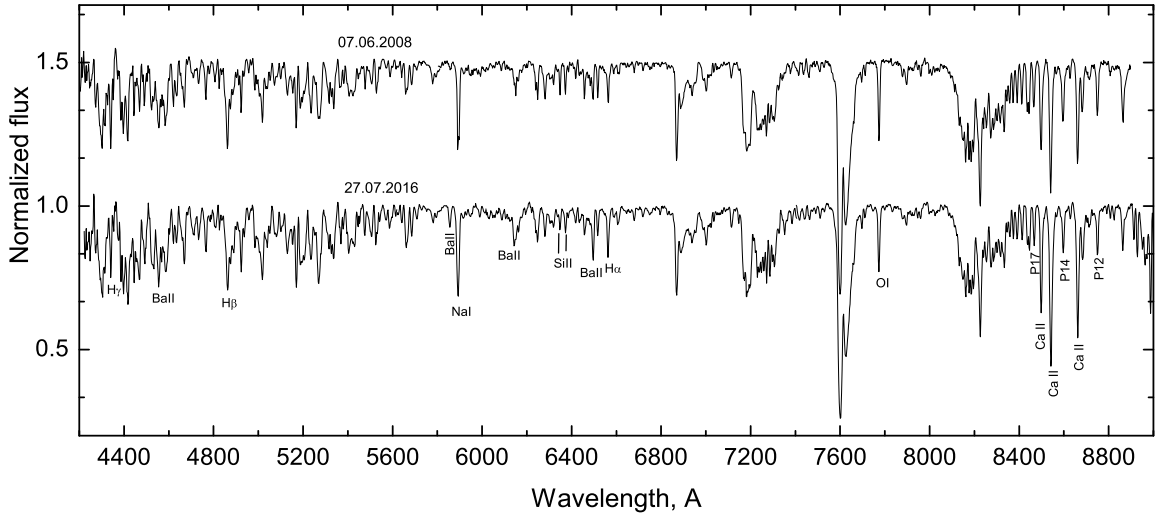}

\caption{The continuum normalized spectra for V1427~Aql obtained
on June 7, 2008, and on July 27, 2016. The upper spectrum is
plotted with a vertical shift of 0.5.}

\end{figure}



\begin{figure}

\includegraphics[scale=1.5]{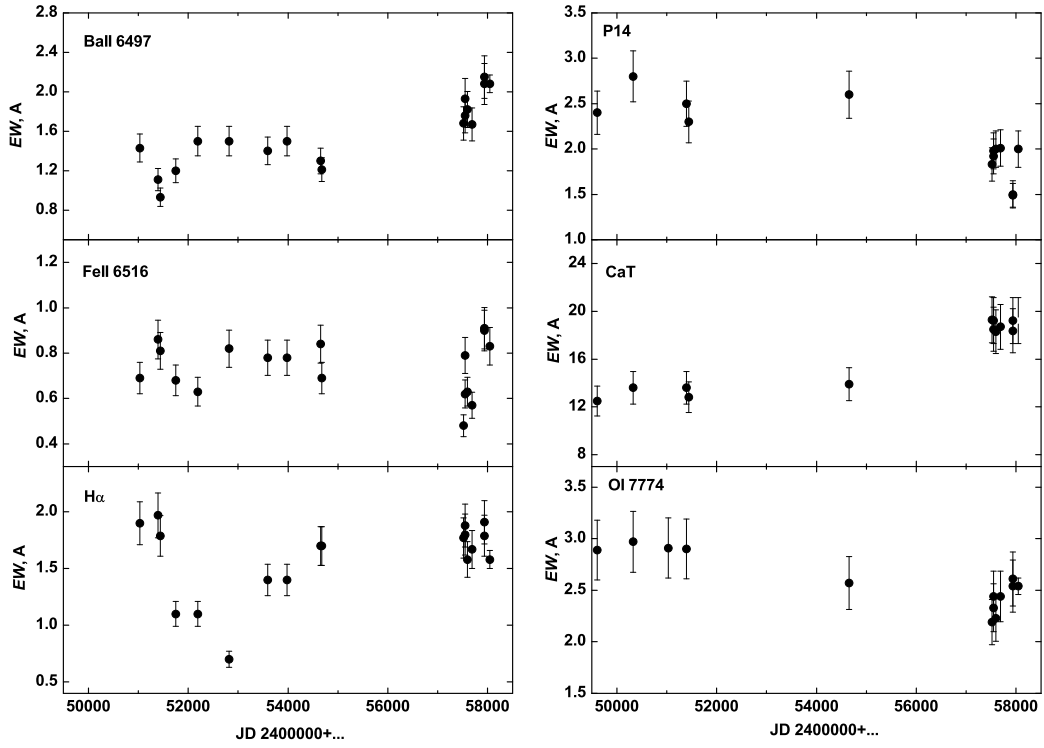}
\caption{Plot showing evolution of the absorption line equivalent widths for V1427~Aql.}

\end{figure}


In our spectra, the H$\alpha$ equivalent width varies more than
twice from 0.7 \AA\ to 1.97 \AA. High-resolution spectra
demonstrate that H$\alpha$ has a complicated profile with varying
emission components on the blue and red sides and with the
absorption feature undergoing significant changes in width and
radial velocity (Tamura and Takeuti 1991; Zacs et al. 1996; Kipper
2008; \c{S}ahin et al. 2016). As stated in Klochkova (2014), mass
loss rate, stellar wind velocity, kinematics and optical depth of
the circumstellar matter are the principal factors that determine
the shape and variability of the H$\alpha$ profile for the far
evolved stars surrounded by spatially extended envelopes.

We can not measure the OI $\lambda$7771-4 triplet components
separately. The OI blend equivalent width was estimated to be
2.9$\pm$0.1 \AA\ in 1994-1999 (an average value), 2.57$\pm$0.30
\AA\ in 2008, and by 2016 it decreased to 2.33$\pm$0.10 \AA. In
literature we could find few $EW$(OI) estimates: 2.6 \AA\ (Reddy
and Hrivnak 1999), 2.9 \AA\ (Pooley 2003), 3.0 \AA\ (S\'{a}nchez
Contreras et al. 2008) -- derived from high-resolution spectra
obtained on October 15-16,1997, May 7, 1999, and June 2, 2003,
respectively.

The oxygen triplet is a good indicator of luminosity for A-G
supergiants. As has been pointed out repeatedly, V1427~Aql has an
extremely high $EW$(OI) value that leads to the absolute magnitude
estimate of $M_{V}\approx-8.^{m}0\pm0.1$ (Reddy and Hrivnak 1999).
It should be kept in mind, however, that the OI triplet strength
also depends on the star effective temperature (Kovtyukh et al.
2012). For example, Pereira and Miranda (2007) demonstrate the
$EW$(OI)-spectral class and $EW$(OI)-luminosity class relations
showing that $EW$(OI) takes its maximum value for F0 supergiants
and monotonically decreases for later spectral classes. Thus we
may assume that the OI triplet weakening detected in the V1427~Aql
spectrum implies the decreasing of stellar temperature since 2008
till 2016.

As could be seen in Table~4, the equivalent widths of Paschen
lines in the V1427~Aql spectrum slightly decreased by 2016-2017,
while the CaII triplet absorptions turned stronger. The total
equivalent width of three lines constituting the CaII triplet was
$EW$(CaT)=12.9 $\pm$0.9 \AA\ in the 1994-2008 interval whereas it
grew to $EW$(CaT)=18.8$\pm$0.4 \AA\ by 2016-2017.

\section*{SPECTRAL CLASS AND EVOLUTION OF STELLAR PARAMETERS}

The spectral class for V1427~Aql was determined more than once.
The HD catalog gives the G5 class. The star was classified as GIa
(Bidelman 1981), G5Ia (Buscombe 1984), G40-Ia (Keenan and McNeil
1989), G5Ia (Hrivnak et al. 1989), F8I (Volk and Kwok 1989),
F8/G0Ia (Houk and Swift 1999) in the eighties-nineties of the last
century. The spectroscopic atlas of post-AGB stars and related
objects (Su\'{a}rez et al. 2006) classifies V1427~Aql as an F7I
star basing on the spectra obtained in June 1995. So, there is a
discrepancy in the star spectral classification in literature. It
is not clear whether it is due to the real change of spectral
class or resulted from the application of different classification
criteria to a peculiar star spectrum.

There were several studies that attempted to evaluate the
atmospheric parameters for V1427~Aql applying the high-resolution
spectra analysis. Zacs et al. (1996) got $T_{eff}$=6800~K, Reddy
and Hrivnak (1999) found the $T_{eff}$ value equal to 6750~K,
Th\'{e}venin et al. (2000) derived $T_{eff}$=5660~K. Kipper et al.
(2008) estimated the effective temperature to be 6750$\pm$200~K.
\c{S}ahin et al. (2016) basing on their observational data
obtained in the 2008-2013 interval derived some of the stellar
parameters including those determined earlier by other
researchers. In \c{S}ahin et al. (2016) the effective temperature
was found higher than previously, $T_{eff}=7350\pm200$~K, and no
systematic change of temperature was detected from 1992 till 2008.
Such a high $T_{eff}$ value leads to about F0 spectral class in
the calibration of Strai\v{z}ys (1977).

Reddy and Hrivnak (1999) pointed out the difference for V1427~Aql
between the spectral class derived from low-resolution spectra and
the effective temperature yielded by the model atmospheres method
operating on high-resolution spectra. New $T_{eff}$ estimate
presented by \c{S}ahin et al. (2016) exacerbates this discrepancy.

We estimated the effective temperature and spectral class for
V1427~Aql using our low-resolution spectra. Table~4 lists the
$T_{eff}$ values derived from the empirical relations between the
ratio of CaII IR-triplet lines equivalent widths to those of
hydrogen Paschen lines (P12, P14, and P17) and the effective
temperature (equation~(1) from Mantegazza (1991)). Our
observational data yield an average $T_{eff}$ value of
6870$\pm$180~K for the 1994-2008 interval. The temperature
decreased to $T_{eff}=5990\pm100$~K by 2016-2017. According to the
Strai\v{z}ys (1977) calibration for supergiants, these
temperatures correspond to F2-F5 spectral classes in 1994-2008 and
about F8 in 2016-2017.

A grid of stellar parameters as a function of CaT* and PaT created
from the Cenarro et al. (2002) study is adduced in a paper that
aims to search for yellow supergiants in the M33 galaxy
(Kourniotis et al. 2017). The PaT index is equal to the sum of
P12, P14, and P17 equivalent widths, whereas the CaT* index
(Cenarro et al. 2001) measures the equivalent width of the
integrated CaII triplet corrected for the contamination from the
adjacent Paschen lines: CaT*=CaT--0.93$\times$PaT. If to plot our
measurements for V1427~Aql in the PaT-CaT* plane, two states can
be distinguished for the star: that with larger values of
$T_{eff}$ and $\log g$ observed in 1994-2008 and the other with
smaller ones taking place in 2016-2017 (Fig.~11).


\begin{figure}

\includegraphics[scale=1.0]{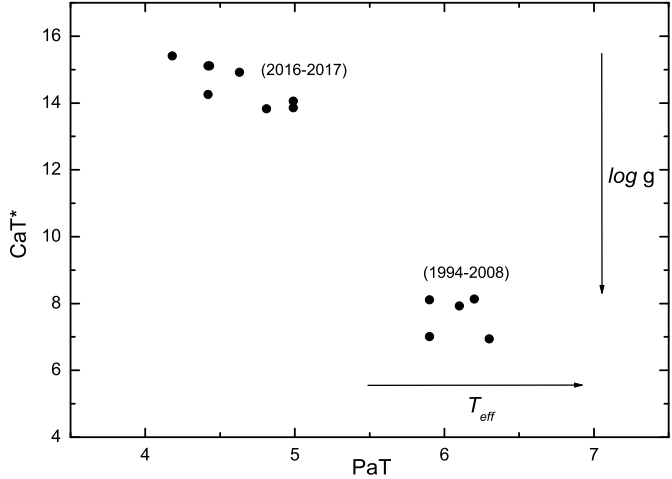}

\caption{PaT-CaT* diagnostic plot for V1427~Aql involving
observations of the 1994-2008 and 2016-2017 intervals. The PaT and
CaT* indices characterize the strength of Paschen lines P12, P14,
P17 and of the CaII IR-triplet absorptions, respectively. Arrows
indicate the direction of stellar parameters, $\log g$ and
$T_{eff}$, growing.}

\end{figure}


We also estimated $T_{eff}$ from our photometric data. For this
purpose the annual average $B-V$ color values corrected for
interstellar reddening with $E(B-V)=0.^{m}7$ were converted to
$T_{eff}$ according to the ($B-V$)--$T_{eff}$ relation for normal
supergiants from Flower (1996). Besides we calculated $T_{eff}$
using the ($V-K$)--$T_{eff}$ relation from Bessel et al. (1998).
Fig.~12 shows the temporary temperature change. $T_{eff}$ displays
a clear tendency to increase since the first observations in 1990
($\sim$ 5500 K) reaching the maximum value ($\sim$ 6000 K) in the
2001-2003 interval and to decrease later on to $\sim$5300 K. What
provokes concern is that the estimates derived from optical
observations start to diverge from those yielded by near-IR data
after the year 2014.

The effective temperature estimated from the optical and near-IR
photometry is slightly lower than that derived from spectral
observations. This difference may be due to the fact that the
"photometric"\ temperature should reflect the state of the
so-called ''pseudo-photosphere'', i.e., of the continuum forming
region in the stellar wind, whereas the "spectral"\ temperature
refers to the absorption forming zone, as well as to the possible
underestimate of interstellar reddening. If $E(B-V)=1.^{m}0$ is
adopted for the object, the "photometric"\ and "spectral"\
temperatures will converge. We should say that interstellar
extinction for HD~179821 has been determined more than once and
the $A_{V}$ value given by different authors varies from 1$^{m}$
to 4$^{m}$ (Pottasch and Parthasarathy 1988; Reddy and Hrivnak
1999; Arkhipova et al. 2001; Arkhipova et al. 2009).


\begin{figure}

\includegraphics[scale=1.5]{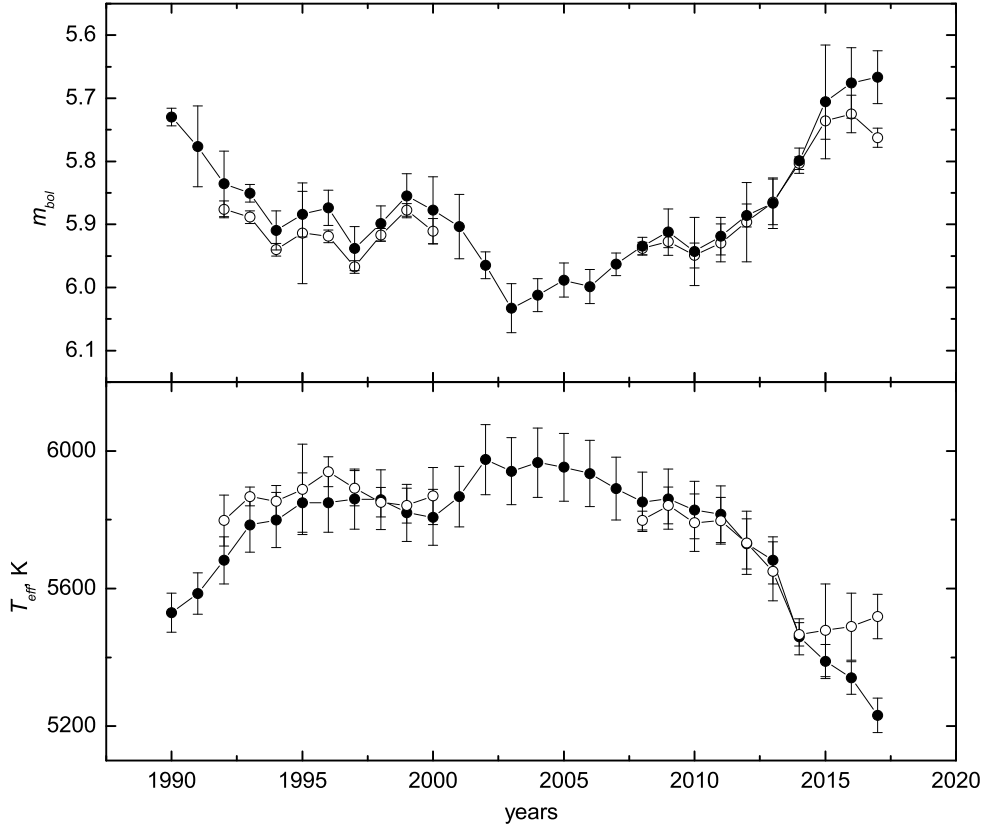}

\caption{Change of effective temperature and bolometric brightness
with time for V1427~Aql. Upper panel: bolometric magnitudes
derived from $m_{bol}=V-A(V)+BC(V)$ are plotted with points, open
circles correspond to $m_{bol}=K-A(K)+BC(K)$. Lower panel: points
connected by a line represent the $T_{eff}$ estimates derived from
the ($B-V$)--$T_{eff}$ for normal supergiants given in Flower
(1996). Open circles correspond to effective temperatures inferred
by the $(V-K)$--$T_{eff}$ relation of Bessel et al. (1998).}

\end{figure}


We have also measured the annual average bolometric brightness for
V1427~Aql $m_{bol}=V-A(V)+BC(V)$ and $m_{bol}=K-A(K)+BC(K)$ using
the $BC(V)$ and $BC(K)$ bolometric corrections from Bessel et al.
(1998) and adopting $A(V)=3.1\times E(B-V)$ and $A(K)=0.28\times
E(B-V)$ according to Koornneef (1983). The upper panel of Fig.~12
displays the $m_{bol}$ change with time. The bolometric magnitude
varies over a range of $0.^{m}3$ that corresponds to a $\sim$30~\%
change in luminosity.

Photometric data evidence that the photosphere temperature
decreased by $\sim$400~K between 2008 and 2016, the luminosity
grew by $\sim19~\%$, and the radius increased by $\sim24~\%$
according to the Stefan-Boltzmann law.

\section*{DISCUSSION}

Owing to the multi-year photometric monitoring of V1427~Aql, we
can conclude that the star displays a semi-periodic variation and
a long-term trend in brightness and color as well. Spectral and
photometric data make it evident that the stellar parameters have
changed in a short time.

Let us try to explain these changes with some assumptions made
about the object's nature.

The star has already passed through a considerable part of its
evolution that is implied by the presence of an expanded nearly
spherical gas-dust circumstellar envelope. V1427~Aql has a
dual-peak energy distribution in the $\lambda\lambda 0.4-100~\mu$m
wavelength range that is characteristic of post-AGB objects
(Pottasch and Parthasarathy 1988; Hrivnak et al. 1989). The
structures that provide the primary input into the energy emitted
by the system are the stellar photosphere and the cold dust
envelope with a temperature of $T_{d}=128$~K (Odenwald 1986).

According to theoretical simulations, the stars with low and
intermediate masses having left the Asymptotic Giant Branch (AGB)
continue to evolve to higher temperatures with the luminosity
staying on a constant level. New photometric and spectral data for
V1427~Aql demonstrate that in recent years a rapid cooling of
photosphere has occurred, the star luminosity has grown, and the
timescale of semi-periodic variation has increased, and all this
does not agree with a model track for a post-AGB object.

A compelling argument for the post-AGB nature of V1427~Aql
consists in a specific abundance pattern derived in Zacs et al.
(1996), Reddy and Hrivnak (1999), Th\'{e}venin et al. (2000),
Kipper (2008). The authors found overabundances of carbon, oxygen,
nitrogen and s-process elements as compared to solar element
distribution, -- a pattern that is typical of post-AGB objects.
But the recent study of \c{S}ahin et al. (2016) states that the
stellar atmosphere is not enhanced in s-process elements and that
the luminosity, effective temperature, and surface gravity of
V1427~Aql are those of a massive star evolving to become a red
supergiant and to explode, in time, as a Supernova II.

Earlier investigations have provided some other marks that
V1427~Aql is a massive object with increased luminosity. For
example, the CO envelope expands with a velocity of
$V_{exp}=33.9$~km/s (Zuckerman and Dyck) that is noticeably larger
than 10-15~km/s that is typical of post-AGB stars. The OI triplet
at $\lambda$7774 \AA\ in the V1427~Aql spectrum is reported to
have an equivalent width in the range of 2.3-3.0~\AA\ according to
different estimates, and this value corresponds to the absolute
magnitude of $M_{bol}<-8.^{m}0$ and the luminosity of
$L>10^{5}L_{\odot}$. Such a luminosity and the derived temperature
place the star into the region on the Hertzsprung-Russell (HR)
diagram occupied by yellow hypergiants (YHGs). The YHG region is
bounded on the side of higher temperatures by a so-called Yellow
Evolutionary Void firstly proposed and described in Nieuwenhuijzen
and de Jager (1995). Episodes of large-scale mass loss occurring
in this phase of evolution make the stars move from red to blue
and vice versa in the HR diagram more than once keeping the
bolometric luminosity nearly constant. After the star evolves off
the RSG and reaches a temperature of about 7000~K, an extensive
mass loss ensues, the star is surrounded by a cool photosphere and
starts to move to the RSG region. The envelope turning optically
thin, the star goes to the blue in the HR diagram (Oudmaijer et
al. 2009). The photometric behavior is seen to be in line with
this scenario.

We would like to compare V1427~Aql with other well-studied
hypergiants: $\rho$ Cas, V509~Cas, V1302~Aql. All these stars
display a significant variation in the spectrum and stellar
parameters.

The most famous object, IRC+10~420 (V1302~Aql), which is a strong
IR-source, was found to have increased effective temperature by
$\sim$1000~K in 20 years since the seventies of the last century
(Oudmaijer et al. 1996) and to have changed its spectral class
from F (Humphreys et al. 1973) to A5 (Oudmaijer et al. 1996;
Klochkova et al. 1997; Klochkova et al. 2002). The stellar
temperature was reported to increase at a rate of 120~K per year
which caused the star to approach the Wolf-Rayet phase (Klochkova
et al. 2002; Humphreys et al. 2002). Continuing to monitor
V1302~Aql in 2001-2014 Klochkova et al. (2016) conclude that the
hypergiant has entered the phase, during which the effective
temperature growth slows down (or ceases), and it is located close
to the high-temperature boundary of the Yellow Void in the HR
diagram.

$\rho$~Cas, a prototype of YHGs, displayed a significant variation
in brightness and spectrum but of another kind than V1302~Aql
(Halbedel 1991). The star is known to have experienced three
episodes of extensive mass loss during the last 100 years: in
1945-1947, in 1985-1986, and in 2000-2001, the latter was
described in detail in Lobel et al. (2003). From the June to
September of 2000 the star brightness decreased by nearly 1$^{m}$,
and the star got out of minimum by the April of 2001. The TiO
bands appeared in the spectrum in the summer of 2000; the spectral
modeling revealed that the temperature had decreased by at least
3000~K, and the mass loss rate during that episode was estimated
to be $\dot{M}\simeq5.4\times10^{-2}M_{\odot}$. The star is
considered to lose mass extremely rapidly in the course of the
so-called ''shell episodes'' and to appear wrapped for several
hundreds of days in a pseudo-photosphere formed by ejected cool
matter. Klochkova et al. (2014) have studied in detail the optical
spectrum and have picked out kinematic properties of the expanded
stellar atmosphere.

The photometric history was monitored for V509~Cas for a hundred
and fifty years (Zsoldos 1986a; Percy and Zsoldos 1992; Halbedel
1993; Nieuwenhuijzen et al. 2012). V509~Cas was observed to change
brightness and color monotonically (Percy and Zsoldos 1992;
Nieuwenhuijzen et al. 2012). Nieuwenhuijzen et al. (2012) showed
that the effective temperature, $\log g$, radius, and luminosity
of the star had changed a lot with time. The $\log g$ and
turbulent velocity $\xi_{t}$ variation argues that the star has
undergone a series of subsequent gas outflows. During every
outflow, the pseudo-photosphere happens to have smaller $\log g$
and larger $\xi_{t}$. After the expelled envelope is dispersed
into space and the stellar atmosphere is adjusted, a more compact
and hot atmosphere is seen.

Let us consider the pulsation activity of the stars
above-mentioned.

There is no data evidencing the periodic light variation for
V1302~Aql; the star shows only monotonic long-term trends in
brightness.

V509~Cas displays semi-periodic variation with a period of
$\sim400^{d}$ and an amplitude up to 0.$^{m}$2 (Zsoldos 1986b).
Percy and Zsoldos (1992) found three peaks in the frequency
spectrum at $P=203.^{d}$3, 299.$^{d}$2, and 385.$^{d}$4 and a
tendency for the fundamental mode amplitude to decrease along with
the $B-V$ color getting bluer.

The variable star $\rho$~Cas, apart from dimming by more than
1$^{m}$ as described above, shows semi-periodic variation with
smaller amplitude. Arellano Ferro (1985), based on 1964-1981
observations, found a period of 483.$^{d}$5.

The object under study, V1427~Aql, displayed a semi-periodic
brightness variation with periods in the range 170-200$^{d}$ and
with the timescale and the amplitude of the oscillation increasing
in recent years. The star undergoes temperature variation as well
as brightness change with no apparent change of color.

Fadeev (2011) in his paper devoted to the simulations of yellow
hypergiants pulsational instability has referred to V1427~Aql as
to an object that can not be a hypergiant because of its too long
variation period that is larger than the uppermost value predicted
by the radial pulsation theory for a yellow hypergiant, on the one
hand, and because of its average effective temperature
$T_{eff}=6750$~K (Kipper 2008) corresponding to the region of
pulsational stability, on the other. On this basis, the author
suggests that V1427~Aql is a post-AGB object.

However, it should be noted that oscillation periods for known
post-AGB stars in the range of spectral classes from F0 to G5 are
not larger than 150$^d$ (Arkhipova et al. 2003; Arkhipova et al.
2010; Hrivnak et al. 2010; Hrivnak et al. 2015). Besides, the true
hypergiants $\rho$~Cas and V509~Cas vary with periods $> 200^{d}$
having temperatures $T_{eff}$ for which radial pulsations are
forbidden by theory. Arellano Ferro (1985) suggested that $\rho$
Cas and V509~Cas are non-radial pulsators. The same mechanism may
appear responsible for the semi-periodic variation of V1427~Aql.

So, analyzing the photometric and spectral variability of
V1427~Aql showed that the star's behavior is similar to that of
yellow hypergiants located near the very unstable Yellow Void and
differs significantly from ordinary post-AGB objects.

\section*{CONCLUSION}

New photometry and spectra obtained for the yellow hypergiant with
a gas-dust envelope, V1427~Aql, and their comparison with archival
data lead us to the following results.

1) Variation periods $P=170^{d}$ and $P=141^{d}$ (or $P=217^{d}$)
were found on the basis of $UBV$-observations in 2009-2011. It is
shown that oscillations became less regular after 2011, the
characteristic time of brightness variation increased to 260$^{d}$
in 2012-2014, and to 370$^{d}$ in the 2015-2017 interval.

2) The annual average brightness was found to gradually increase
in 2009-2015 in the $V$, $J$, $K$ filters, and to decrease in $U$
and $B$. The annual average $U-B$, $B-V$, $J-K$ colors were
growing (the star was getting redder). The near-IR brightness
stopped increasing in 2016-2017, and V1427~Aql showed a tendency
to get bluer, whereas it kept on getting redder in the optical.

3) The long-term trend in colors was proved to reflect the change
of photosphere temperature, increasing since 1990 till 2002 and
then systematically decreasing. Long-term variation of the
bolometric magnitude does not agree with the assumption of quick
evolution at constant luminosity but seems to be related to mass
loss episodes that result in the expanded stellar envelope getting
optically thicker and the continuum spectrum originating in more
outer and cool layers.

4) The comparison of the 2016-2017 spectral data for V1427~Aql
with those obtained in 1994-2008 revealed the strengthening of
BaII and CaII IR-triplet absorptions and the weakening of OI
triplet blend. These changes are supposed to be due to the
decreasing of temperature in the region where the absorptions are
formed.

It is very important to continue photometric and spectral
observations of the object that changes so quickly. It is highly
recommended to carry out both photometric and spectral monitoring
of the star with no durable gaps.

 \bigskip

ACKNOWLEDGMENT

This study has made use of the SIMBAD and VIZIER databases and the NASA ADS.

 \bigskip

REFERENCES

\begin{enumerate}

\item A. Arellano Ferro, MNRAS {\bf 216}, 571 (1985).

\item V.P. Arkhipova, N.P. Ikonnikova, R.I. Noskova, Astron Lett.
{\bf 19}, 169 (1993).

\item V.P. Arkhipova, N.P. Ikonnikova, R.I. Noskova, G.V. Sokol,
S.Yu. Shugarov, Astron. Lett. {\bf 27}, 156 (2001).

\item  V.P. Arkhipova, R.I. Noskova, N.P. Ikonnikova, G.V. Komissarova,
    Astron. Lett. {\bf 29}, 480 (2003).

\item V.P. Arkhipova, V.P.Esipov, N.P. Ikonnikova, G.V.
Komissarova, A.M. Tatarnikov, B.F. Yudin, Astron. Lett. {\bf 35},
764 (2009).

\item V.P. Arkhipova, N.P. Ikonnikova, G.V. Komissarova, Astron.
Lett. {\bf 36}, 269 (2010).

\item M.S. Bessell, F. Castelli, P. Planesas, Astron. Astrophys.
{\bf 333}, 231 (1998).

\item W.P.Bidelman, Astron. J. {\bf 86}, 553 (1981).

\item W. Buscombe, MK Spectral Classification, Sixth General
Catalog (Evanston: Northwestern University, 1984).

\item A.J. Cenarro, P.N. Cardiel, J.Gorgas, R.F. Peletier,
A.Vazdekis, and F. Prada, MNRAS {\bf 326}, 959 (2001).

\item A. J. Cenarro, J. Gorgas, N. Cardiel, A. Vazdekis, and R.F.
Peletier, MNRAS {\bf 329}, 863 (2002).

\item H. Coroller, A. L\'{e}bre, D. Gillet, and E. Chapellier,
Astron. Astrophys. {\bf 400}, 613 (2003).

\item T. J. Deeming, Astrophys. and Space Sci. {\bf 36}, 137 (1975).

\item Yu.A. Fadeev, Astron. Lett. {\bf 37}, 403 (2011).

\item B.A. Ferguson and T. Ueta, Astrophys. J. {\bf 711}, 613 (2010).

\item P.J. Flower, Astrophys. J. {\bf 469}, 355 (1996).

\item I.N. Glushneva, V.T. Doroshenko, T.S. Fetisova, T.S.
Khruzina, E.A. Kolotilov, L.V. Mossakovskaya, S.L. Ovchinnikov,
and I.B.Voloshina, VizieR Online Data Catalog III/208 (1998).

\item E.M. Halbedel, IBVS No. 3616, 1 (1991).

\item E.M. Halbedel, IBVS No. 3849, 1 (1993).

\item N. Houk and C. Swift, {\it Michigan catalogue of
two-dimensional spectral types for the HD Stars} vol. 5. By Nancy
Houk and Carrie Swift (Ann Arbor, Michigan: Department of
Astronomy, University of Michigan, 1999).

\item B.J. Hrivnak, S. Kwok, and  K.M. Volk, Astrophys. J. {\bf
346}, 265, (1989).

\item B.J. Hrivnak, in {\it Post-AGB Objects as a Phase of Stellar
Evolution, Proceedings of the Torun Workshop held July5-7, 2000}.
Ed. R. Szczerba and S.K. G\'{o}rny, Astroph. Space Sci. Library
{\bf 265} (Kluwer Academic Pub, 2001), p. 101

\item B.J. Hrivnak, W. Lu, R.E. Maupin, and B.D. Spitzbart,
Astrophys. J. {\bf 709}, 1042 (2010).

\item B.J. Hrivnak, W. Lu, and K. A. Nault, Astron. J. {\bf 149},
184 (2015).

\item R.M. Humphreys, D.W. Strecker, T.L. Urdock, and F.J. Low,
Astrophys. J. {\bf 179}, L49-L52 (1973).

\item R.M. Humphreys, K. Davidson, and N. Smith, Astron. J. {\bf
1724}, 1026 (2002).

\item H.L. Johnson, R.I. Mitchel, B. Iriarte, and W.Z. Wisniewski,
Comm. Lunar and Planet. Lab. {\bf4}, 99 (1966).

\item E. Josselin and A. Lebre, Astron. Astrophys. {\bf 367}, 826
(2001).

\item M. Jura, T. Velusamy, and M.W. Werner, Astrophys. J. {\bf
556}, 408 (2001).

\item J.H. Kastner and D.A. Weintraub, Astrophys. J. {\bf 452},
833 (1995).

\item P.C. Keenan and R.C.  McNeil, Astrophys. J. Suppl. Ser. {\bf
71}, 245 (1989).

\item T. Kipper, Baltic Astronomy {\bf 17}, 87 (2008).

\item V. Klochkova,  E.L. Chentsov, and V. Panchuk, MNRAS {\bf
292}, 19 (1997).

\item V.G. Klochlova, M.V. Yushkin, E.L. Chentsov, and V.E.
Panchuk, Astron. Rep. {\bf 46}, 139 (2002).

\item V.G. Klochkova, Astrophys. Bull. {\bf 69}, 279 (2014).

\item V.G. Klochlova, V.E. Panchuk, N.S. Tavolzhanskaya, and I.A.
Usenko, Astron. Rep. {\bf 58}, 101 (2014).

\item V.G. Klochkova, E.L. Chentsov, A.S. Miroshnichenko, V.E.
Panchuk, and M.V. Yushkin, MNRAS {\bf 459}, 4183 (2016).

\item J. Koornneef, Astron. Asrophys. {\bf 128}, 84 (1983).

\item M. Kourniotis, A. Z. Bonanos, W. Yuan, L. M. Macri, D.
Garcia-Alvarez, and C.-H. Lee, Astron. Astrophys. {\bf 601},
id.A76, 21 (2017).

\item V. V. Kovtyukh, N. I. Gorlova, and S. I. Belik,  MNRAS {\bf
423}, 3268 (2012).

\item P. Lenz and M. Breger, Communications in Asteroseismology,
{\bf 146}, 53 (2005).

\item A. Lobel, A. K. Dupree, R. P. Stefanik, G. Torres, G.
Israelian, N. Morrison, C. de Jager,  H. Nieuwenhuijzen, I. Ilyin,
and F. Musaev, Astrophys. J. {\bf 583}, 923 (2003).

\item V. M. Lyutyi, Soobshch. GAISh {\bf 172}, 30 (1971).

\item L. Mantegazza, Astron. Astrophys. Suppl. {\bf 88}, 255 (1991).

\item H. Nieuwenhuijzen and C. de Jager, Astron. Astrophys. {\bf
302}, 811 (1995).

\item H. Nieuwenhuijzen, C. De Jager, I. Kolka, G. Israelian, A.
Lobel, E. Zsoldos, A. Maeder, and G. Meynet, Astron. Astrophys.
{\bf 546}, A105 (2012).

\item S.F. Odenwald, Astrophys. J. {\bf 307}, 714 (1986).

\item R.D. Oudmaijer, M.A.T. Groenewegen, H.E. Matthews, J.A.D.L.
Blommaert, and K.C. Sahu, MNRAS {\bf 280}, 1062 (1996).

\item R.D. Oudmaijer, B.Davies, W.-J. de Wit, and M. Patel, in
{\it The Biggest, Baddest, Coolest Stars, ASP Conference Series,
Vol. 412, Proceedings of the workshop held 16-18 July 2007, at the
Millennium Centre, Johnson City, Tennessee, USA}. Ed. by D.G.
Luttermoser, B.J. Smith, and R.E. Stencel (San Francisco:
Astronomical Society of the Pacific, 2009), p.17.

\item J.R. Percy and E. Zsoldos, Astron. Astrophys., {\bf 263},
123 (1992).

\item C.B. Pereira and L.F. Miranda, Astron. Astrophys. {\bf 462},
231 (2007).

\item A.J. Pickles, PASP  {\bf 110}, 863 (1998).

\item D. Pooley, Thesis, University of Canterbury (2003).

\item S.R. Pottasch and М. Parthasarathy, Astron. Astrophys. {\bf
192}, 182 (1988).

\item B.E. Reddy and B.J. Hrivnak, Astron. J. {\bf 117}, 1834
(1999).

\item T. \c{S}ahin, D.L. Lambert, V.G. Klochkova, and V.E.
Panchuk, MNRAS {\bf 461}, 4071 (2016).

\item N.N. Samus, E.V. Kazarovets, O.V. Durlevich, N.N. Kireeva,
and E.N. Pastukhova, Astron. Rep. {\bf 61}, 80 (2017).

\item C. S\'{a}nchez Contreras, R. Sahai, A. Gil de Paz, and R.
Goodrich, Astrophys. J. Suppl. Ser. {\bf 179}, 166 (2008).

\item S.G. Sergeev and F. Heisberger, A Users Manual for SPE. Wien
(1993).

\item V.I. Shenavrin, O.G. Taranova, and A. E. Nadzhip, Astron,
Rep. {\bf  55}, 31 (2011).

\item V.L. Strai\v{z}ys, {\it Mnogotsvetnaya fotometriya zvezd}
(Vilnius: Mokslas, 1977).

\item O. Suarez, J.F. Gomez, and O. Morata, Astron. and Astrophys.
{\bf 458}, 173 (2006).

\item S. Tamura, M. Takeuti, IBVS No. 3561 (1991).

\item F. Th\'{e}venin, M. Parthasarathy, and G. Jasniewicz,
Astron. Astrophys. {\bf 359}, 138 (2000).

\item W.E.C.J. van der Veen, H.J. Habing and T.R. Geballe, Astron.
Astrophys. {\bf 226}, 108 (1989).

\item F. van Leeuwen, Astron. Astrophys. {\bf 474}, 653 (2007).

\item K.M. Volk and S. Kwok, Astrophys. J. {\bf 342}, 345 (1989).

\item L. Zacs, V.G. Klochkova, V.E. Panchuk, and R. Spelmanis,
MNRAS {\bf 282}, 1171 (1996).

\item E. Zsoldos, The Observatory {\bf 106}, 156 (1986a).

\item E. Zsoldos, in {\it Luminous Stars and Associations in
Galaxies, Proceedings of the IAU Symposium No. 116}. Ed. by de
Loore C.W.H., Willis A.J., Laskarides P. (Reidel Publ. Co.,
Dordrecht, 1986b), p. 87.

\item B. Zuckerman and H.M. Dyck, Astrophys. J. {\bf 311}, 345
(1986).

\end{enumerate}

\newpage

\begin{center}
\begin{longtable}{cccccc}
\caption{$UBV$-photometry for V1427 Aql in 2009-2017.}\\
\hline
\endfirsthead

\multicolumn{4}{c}{continued \tablename\ \thetable{}}\\
\hline
JD&   $V$ &  $B$ &  $U$ &  $B-V$ & $U-B$\\
\hline
\endhead

\hline
\endlastfoot

\hline
JD&   $V$ &  $B$ &  $U$ &  $B-V$ & $U-B$\\
\hline

2454970&   8.030&   9.331&   10.453&   1.301&   1.122\\
2454972&   8.021&   9.345&   10.461&   1.323&   1.116\\
2454978&   8.022&   9.331&   10.435&   1.309&   1.105\\
2454981&   8.030&   9.341&   10.387&   1.310&   1.046\\
2454983&   8.048&   9.343&   10.361&   1.296&   1.018\\
2455002&   8.064&   9.383&   10.532&   1.319&   1.149\\
2455005&   8.054&   9.358&   10.432&   1.305&   1.073\\
2455010&   8.055&   9.364&   10.490&   1.309&   1.126\\
2455031&   8.093&   9.467&   10.557&   1.374&   1.089\\
2455032&   8.113&   9.427&   10.574&   1.314&   1.147\\
2455036&   8.096&   9.416&   10.591&   1.320&   1.175\\
2455041&   8.097&   9.416&   10.556&   1.319&   1.140\\
2455043&   8.067&   9.431&   10.635&   1.364&   1.204\\
2455055&   8.063&   9.369&   10.516&   1.306&   1.147\\
2455061&   8.058&   9.369&   10.486&   1.310&   1.118\\
2455065&   8.052&   9.363&   10.448&   1.310&   1.085\\
2455069&   8.032&   9.345&   10.446&   1.312&   1.101\\
2455071&   8.032&   9.335&   10.445&   1.303&   1.110\\
2455073&   8.046&   9.345&   10.459&   1.299&   1.114\\
2455086&   8.004&   9.287&   10.374&   1.283&   1.087\\
2455090&   7.996&   9.267&   10.331&   1.271&   1.065\\
2455093&   7.985&   9.274&   10.314&   1.289&   1.040\\
2455098&   7.983&   9.263&   10.372&   1.280&   1.109\\
2455103&   8.003&   9.281&   10.399&   1.278&   1.118\\
2455112&   7.993&   9.287&   10.384&   1.295&   1.097\\
2455113&   7.986&   9.279&   10.393&   1.293&   1.114\\
2455118&   8.007&   9.268&   10.381&   1.261&   1.113\\
2455145&   8.014&   9.298&   10.381&   1.284&   1.083\\
2455151&   8.014&   9.318&   10.368&   1.305&   1.049\\
2455159&   8.068&   9.355&   10.390&   1.287&   1.034\\
2455164&   8.066&   9.370&   10.450&   1.304&   1.081\\
2455281&   8.021&   9.325&   10.419&   1.305&   1.094\\
2455291&   8.037&   9.337&   10.458&   1.300&   1.121\\
2455305&   8.080&   9.391&   10.526&   1.310&   1.135\\
2455313&   8.079&   9.403&   10.546&   1.323&   1.143\\
2455315&   8.092&   9.422&   10.619&   1.330&   1.198\\
2455329&   8.121&   9.470&   10.666&   1.349&   1.196\\
2455330&   8.137&   9.475&   10.641&   1.338&   1.166\\
2455341&   8.121&   9.468&   10.642&   1.347&   1.174\\
2455344&   8.144&   9.479&   10.655&   1.335&   1.176\\
2455349&   8.117&   9.479&   10.645&   1.362&   1.165\\
2455357&   8.109&   9.447&   10.553&   1.338&   1.106\\
2455358&   8.086&   9.439&   10.576&   1.353&   1.137\\
2455364&   8.115&   9.443&   10.610&   1.328&   1.167\\
2455369&   8.092&   9.411&   10.573&   1.319&   1.162\\
2455380&   8.075&   9.399&   10.499&   1.323&   1.100\\
2455389&   8.042&   9.345&   10.494&   1.303&   1.149\\
2455408&   7.989&   9.277&   10.391&   1.288&   1.114\\
2455418&   7.987&   9.259&   10.358&   1.272&   1.099\\
2455424&   7.979&   9.258&   10.335&   1.279&   1.076\\
2455445&   8.001&   9.290&   10.391&   1.289&   1.101\\
2455448&   8.009&   9.288&   10.423&   1.279&   1.135\\
2455454&   8.019&   9.318&   10.456&   1.299&   1.138\\
2455463&   8.035&   9.335&   10.494&   1.300&   1.159\\
2455468&   8.054&   9.357&   10.488&   1.303&   1.132\\
2455492&   8.108&   9.440&   10.622&   1.333&   1.181\\
2455494&   8.118&   9.454&   10.632&   1.336&   1.178\\
2455501&   8.131&   9.491&   10.662&   1.359&   1.172\\
2455507&   8.137&   9.482&   10.681&   1.346&   1.199\\
2455516&   8.125&   9.457&   10.617&   1.332&   1.161\\
2455678&   8.072&   9.403&   10.540&   1.331&   1.137\\
2455688&   8.074&   9.406&   10.576&   1.333&   1.169\\
2455692&   8.074&   9.404&   10.572&   1.330&   1.168\\
2455697&   8.074&   9.404&   10.562&   1.330&   1.159\\
2455708&   8.072&   9.403&   10.540&   1.331&   1.137\\
2455715&   8.074&   9.406&   10.576&   1.333&   1.169\\
2455720&   8.074&   9.404&   10.572&   1.330&   1.168\\
2455734&   8.074&   9.404&   10.562&   1.330&   1.159\\
2455737&   8.076&   9.397&   10.573&   1.321&   1.176\\
2455746&   8.083&   9.407&   10.568&   1.323&   1.162\\
2455752&   8.076&   9.401&   10.551&   1.325&   1.150\\
2455763&   8.060&   9.376&   10.525&   1.316&   1.149\\
2455765&   8.072&   9.392&   10.560&   1.320&   1.168\\
2455768&   8.046&   9.378&   10.517&   1.332&   1.139\\
2455774&   8.047&   9.364&   10.496&   1.317&   1.132\\
2455780&   8.033&   9.348&   10.518&   1.315&   1.169\\
2455782&   8.031&   9.344&   10.521&   1.312&   1.177\\
2455795&   8.040&   9.356&   10.521&   1.316&   1.165\\
2455810&   8.009&   9.332&   10.480&   1.322&   1.149\\
2455823&   7.995&   9.308&   10.488&   1.312&   1.180\\
2455832&   7.989&   9.300&   10.465&   1.310&   1.165\\
2455841&   8.011&   9.327&   10.513&   1.316&   1.186\\
2455861&   8.087&   9.464&   10.591&   1.377&   1.127\\
2455863&   8.044&   9.379&   10.550&   1.335&   1.170\\
2455866&   8.061&   9.396&   10.557&   1.335&   1.161\\
2455873&   8.077&   9.422&   10.608&   1.346&   1.186\\
2455900&   8.105&   9.429&   10.561&   1.323&   1.133\\
2456040&   8.038&   9.369&   10.529&   1.331&   1.161\\
2456043&   8.037&   9.375&   10.505&   1.338&   1.130\\
2456046&   8.043&   9.402&   10.572&   1.358&   1.170\\
2456051&   8.048&   9.405&   10.592&   1.357&   1.187\\
2456059&   8.031&   9.398&   10.602&   1.367&   1.204\\
2456066&   8.048&   9.400&   10.582&   1.352&   1.182\\
2456093&   8.069&   9.411&   10.624&   1.342&   1.214\\
2456098&   8.062&   9.416&   10.629&   1.354&   1.213\\
2456103&   8.050&   9.411&   10.632&   1.360&   1.221\\
2456118&   8.042&   9.405&   10.594&   1.363&   1.189\\
2456123&   8.041&   9.415&   10.620&   1.374&   1.205\\
2456124&   8.040&   9.418&   10.616&   1.378&   1.199\\
2456128&   8.047&   9.416&   10.625&   1.369&   1.209\\
2456131&   8.042&   9.426&   10.623&   1.384&   1.196\\
2456151&   8.053&   9.433&   10.650&   1.380&   1.217\\
2456156&   8.048&   9.437&   10.654&   1.389&   1.217\\
2456162&   8.041&   9.431&   10.659&   1.391&   1.228\\
2456176&   8.029&   9.402&   10.608&   1.373&   1.206\\
2456178&   8.032&   9.396&   10.626&   1.364&   1.230\\
2456202&   8.022&   9.399&   10.633&   1.377&   1.234\\
2456224&   8.003&   9.379&   10.578&   1.376&   1.199\\
2456249&   7.987&   9.338&   10.515&   1.351&   1.132\\
2456423&   8.089&   9.514&   10.849&   1.425&   1.335\\
2456434&   8.088&   9.487&   10.776&   1.399&   1.289\\
2456448&   8.009&   9.398&   10.662&   1.389&   1.265\\
2456464&   7.992&   9.380&   10.601&   1.388&   1.221\\
2456472&   8.020&   9.405&   10.625&   1.385&   1.220\\
2456482&   8.018&   9.404&   10.633&   1.386&   1.229\\
2456487&   8.004&   9.391&   10.629&   1.387&   1.239\\
2456511&   8.012&   9.388&   10.598&   1.376&   1.210\\
2456517&   7.994&   9.358&   10.573&   1.364&   1.215\\
2456519&   7.985&   9.365&   10.562&   1.380&   1.198\\
2456531&   7.985&   9.365&   10.577&   1.380&   1.213\\
2456573&   8.043&   9.458&   10.734&   1.415&   1.276\\
2456575&   8.043&   9.455&   10.762&   1.412&   1.307\\
2456605&   8.050&   9.475&   10.774&   1.425&   1.299\\
2456748&   7.993&   9.401&   10.639&   1.407&   1.239\\
2456780&   8.022&   9.447&   10.772&   1.425&   1.325\\
2456832&   8.026&   9.516&   10.876&   1.490&   1.360\\
2456839&   8.027&   9.553&   10.897&   1.525&   1.345\\
2456869&   8.023&   9.518&   10.896&   1.495&   1.378\\
2456884&   8.028&   9.574&   10.951&   1.547&   1.377\\
2456892&   8.034&   9.540&   10.945&   1.506&   1.405\\
2456902&   8.040&   9.538&   10.929&   1.499&   1.391\\
2456917&   8.020&   9.501&   10.841&   1.481&   1.340\\
2456942&   7.963&   9.380&   10.642&   1.417&   1.262\\
2456960&   7.940&   9.339&   10.583&   1.399&   1.244\\
2457162&   8.047&   9.580&   11.018&   1.533&   1.439\\
2457190&   8.024&   9.544&   11.021&   1.520&   1.477\\
2457213&   8.004&   9.531&   10.982&   1.527&   1.450\\
2457219&   7.998&   9.532&   10.983&   1.534&   1.452\\
2457225&   7.978&   9.507&   10.948&   1.529&   1.441\\
2457232&   7.974&   9.497&   10.956&   1.523&   1.459\\
2457249&   7.913&   9.436&   10.869&   1.523&   1.433\\
2457269&   7.844&   9.324&   10.781&   1.480&   1.457\\
2457299&   7.797&   9.255&   10.612&   1.457&   1.357\\
2457486&   8.012&   9.602&   11.163&   1.590&   1.573\\
2457495&   8.012&   9.588&   11.163&   1.576&   1.575\\
2457516&   8.022&   9.614&   11.204&   1.593&   1.590\\
2457517&   8.038&   9.616&   11.186&   1.578&   1.570\\
2457525&   7.984&   9.575&   11.152&   1.591&   1.577\\
2457544&   7.949&   9.501&   11.016&   1.551&   1.515\\
2457545&   7.950&   9.524&   11.047&   1.574&   1.523\\
2457549&   7.955&   9.499&   11.015&   1.545&   1.515\\
2457556&   7.928&   9.456&   10.979&   1.528&   1.523\\
2457565&   7.909&   9.457&   10.973&   1.549&   1.516\\
2457573&   7.909&   9.451&   10.943&   1.527&   1.469\\
2457575&   7.924&   9.454&   10.965&   1.519&   1.488\\
2457578&   7.926&   9.453&   10.950&   1.509&   1.483\\
2457581&   7.926&   9.452&   10.929&   1.549&   1.413\\
2457597&   7.914&   9.441&   10.910&   1.486&   1.456\\
2457598&   7.919&   9.438&   10.926&   1.471&   1.448\\
2457602&   7.896&   9.405&   10.888&   1.481&   1.443\\
2457609&   7.873&   9.421&   10.834&   1.485&   1.465\\
2457613&   7.872&   9.358&   10.813&   1.499&   1.462\\
2457625&   7.886&   9.357&   10.805&   1.471&   1.448\\
2457634&   7.869&   9.350&   10.793&   1.481&   1.443\\
2457662&   7.897&   9.382&   10.846&   1.485&   1.464\\
2457665&   7.897&   9.395&   10.858&   1.498&   1.462\\
2457870&   8.036&   9.605&   11.143&   1.569&   1.538\\
2457871&   8.028&   9.612&   11.170&   1.584&   1.557\\
2457905&   8.016&   9.603&   11.152&   1.587&   1.549\\
2457931&   8.009&   9.583&   11.120&   1.574&   1.537\\
2457935&   8.001&   9.577&   11.139&   1.576&   1.562\\
2457950&   7.993&   9.576&   11.135&   1.584&   1.559\\
2457955&   7.996&   9.584&   11.157&   1.588&   1.573\\
2457958&   7.983&   9.529&   11.149&   1.547&   1.620\\
2457961&   7.998&   9.591&   11.180&   1.594&   1.589\\
2457966&   7.984&   9.560&   11.136&   1.576&   1.576\\
2457968&   7.977&   9.577&   11.157&   1.599&   1.580\\
2457977&   7.971&   9.529&   11.132&   1.558&   1.603\\
2457979&   7.964&   9.538&   11.111&   1.574&   1.573\\
2457985&   7.968&   9.524&   11.099&   1.556&   1.575\\
2457990&   7.941&   9.506&   11.069&   1.565&   1.563\\
2458006&   7.923&   9.464&   10.994&   1.541&   1.530\\
2458009&   7.917&   9.481&   11.051&   1.564&   1.570\\
2458010&   7.915&   9.480&   11.030&   1.565&   1.550\\
2458015&   7.907&   9.461&   11.015&   1.554&   1.554\\
2458026&   7.914&   9.465&   10.995&   1.551&   1.530\\
2458044&   7.941&   9.502&   11.030&   1.561&   1.528\\
2458075&   7.981&   9.566&   11.175&   1.585&   1.608\\

\end{longtable}
\end{center}

\begin{table}
\begin{center}
\caption{$JHKLM$-photometry for V1427~Aql in 2009-2017.}

\begin{tabular}{cccccc}
\hline
JD&   $J$ &  $H$ &  $K$ &  $L$ & $M$\\
\hline

2455048.35&5.43&5.00&4.75&4.48&--\\
2455319.58&5.43&5.02&4.76&4.54&--\\
2455410.34&5.36&--&4.72&--&--\\
2455433.31&5.39&--&4.73&--&--\\
2455501.15&5.45&--&4.76&--&--\\
2455700.53&5.40&--&4.72&--&--\\
2455734.45&5.44&--&4.74&--&--\\
2455753.40&5.44&--&4.76&--&--\\
2455782.33&5.40&5.00&4.73&4.50&--\\
2455818.28&5.36&--&4.68&--&--\\
2456084.53&5.37&--&4.70&--&--\\
2456147.35&5.31&--&4.61&--&--\\
2456409.58&5.33&--&4.63&--&--\\
2456470.44&5.28&--&4.57&--&--\\
2456489.38&5.28&--&4.57&--&--\\
2456515.39&5.25&--&4.57&--&--\\
2456591.16&5.22&--&4.53&--&--\\
2456876.31&5.13&--&4.41&--&--\\
2456968.19&5.13&--&4.40&--&--\\
2457233.40&5.09&--&4.38&--&--\\
2457261.32&5.08&--&4.35&--&--\\
2457266.30&5.06&4.63&4.34&4.10&3.83\\
2457292.21&5.03&4.60&4.31&4.09&3.70\\
2457525.50&5.10&4.64&4.37&4.11&4.10\\
2457529.54&5.11&4.66&4.39&4.14&4.09\\
2457556.48&5.09&4.62&4.35&4.12&4.11\\
2457566.44&5.08&4.62&4.32&4.07&3.99\\
2457583.38&5.08&4.64&4.35&4.09&4.02\\
2457594.36&5.10&4.66&4.36&4.11&4.02\\
2457616.32&5.04&4.61&4.31&4.10&3.93\\
2457640.32&5.04&4.60&4.30&4.11&3.94\\
2457855.60&5.11&4.64&4.38&4.14&4.14\\
2457892.50&5.10&4.64&4.41&4.16&4.22\\
2457909.50&5.07&4.57&4.30&4.06&3.94\\
2457912.50&5.13&4.63&4.36&4.09&3.95\\
2457943.40&5.08&4.61&4.37&4.06&4.01\\
2457948.40&5.07&4.63&4.35&4.07&4.03\\
2457950.40&5.09&4.61&4.35&4.06&3.97\\
2457971.40&5.07&4.61&4.33&4.08&4.09\\
2457975.30&5.06&4.59&4.33&4.07&4.02\\
2457978.30&5.06&4.61&4.33&4.07&4.13\\
2457999.30&5.05&4.60&4.34&4.10&3.99\\
2458008.30&5.06&4.59&4.30&4.04&4.05\\
2458063.20&5.12&4.63&4.30&4.06&4.00\\

\hline
\end{tabular}
\end{center}
\end{table}

\begin{table}
\caption{Log of spectral observations for V1427~Aql in 2016-2017.}
\begin{center}
\begin{tabular}{ccc}
\hline
 Date & JD& Standard \\

\hline
07.05.2016&2457516&4 Aql\\
05.06.2016&2457545&50 Boo\\
08.06.2016&2457548&4 Aql\\
27.07.2016&2457597&4 Aql\\
25.10.2016&2457687&--\\
27.06.2017&2457932&4 Aql\\
30.06.2017&2457935&4 Aql\\
16.10.2017&2458043&4 Aql\\
\hline
\end{tabular}
\end{center}
\end{table}

\begin{table}

\caption{Equivalent widths for V1427~Aql and the star effective
temperature estimates.}

{\small
\begin{tabular}{cccccccccccc}
\hline

Date &JD&\multicolumn{9}{c} {$EW$, \AA}&$T_{eff}$, K \\
&&H$\alpha$&FeII(6516)&BaII(6497)
&BaII(5854)&OI(7774)&P12&P14&P17&CaT&\\

\hline

12.09.1994&2449608&--&--&--&--&2.9&2.0&2.4&1.5&12.5&6835\\
30.08.1996&2450326&--&--&--&--&3.0&1.8&2.8&1.3&13.6&6720\\
03.08.1998&2451029&1.90&0.70&1.40&0.60&2.9&--&--&--&--&--\\
07.08.1999&2451398&1.97&0.86&1.11&0.29&2.9&2.0&2.5&1.6&13.6&6766\\
20.09.1999&2451442&1.79&0.81&0.93&0.40&--&2.3&2.3&1.7&12.8&6893\\
25.07.2000&2451751&1.10&0.68&1.20&0.46&--&--&--&--&--&--\\
11.10.2001&2452194&1.10&0.63&1.50&0.33&--&--&--&--&--&--\\
02.07.2003&2452823&0.70&0.82&1.50&--&--&--&--&--&--&--\\
09.08.2005&2453592&1.40&0.78&1.40&--&--&--&--&--&--&--\\
30.08.2006&2453978&1.40&0.78&1.50&0.31&--&--&--&--&--&--\\
06.07.2008&2454654&1.70&0.84&1.30&0.24&2.6&2.0&2.6&1.6&13.9&6759\\
01.08.2008&2454680&1.70&0.69&1.21&0.18&--&--&--&--&--&--\\
07.05.2016&2457516&1.77&0.48&1.68&0.74&2.2&1.2&1.8&1.2&19.3&5850\\
05.06.2016&2457545&1.88&0.79&1.93&0.60&2.3&1.8&2.0&1.2&18.5&6124\\
08.06.2016&2457548&1.80&0.62&1.76&0.84&2.4&1.4&1.9&1.1&19.2&5922\\
27.07.2016&2457597&1.58&0.63&1.82&0.80&2.2&1.7&2.0&1.1&18.3&6080\\
25.10.2016&2457687&1.67&0.57&1.67&0.58&2.4&1.9&2.0&1.0&18.7&6100\\
27.06.2017&2457932&1.90&0.91&2.15&0.56&2.6&1.8&1.5&1.1&19.2&5920\\
30.06.2017&2457935&1.79&0.90&2.08&0.54&2.5&1.8&1.5&1.1&18.4&5924\\
16.10.2017&2458043&1.58&0.83&2.08&0.77&2.5&1.7&2.0&0.9&19.2&5975\\

\hline
\end{tabular}
}
\end{table}

 \end{document}